\documentclass[letterpaper, 10 pt, conference]{ieeeconf}

\IEEEoverridecommandlockouts                              

\let\labelindent\relax
\usepackage{graphicx}
\usepackage{subcaption}
\usepackage{amsmath}
\usepackage{amssymb}
\usepackage{accents}
\usepackage{statex}
\usepackage{cases}
\usepackage{tikz,pgfplots}
\pgfplotsset{compat=newest}
\usepackage{xcolor}
\usepackage{caption}
\usepackage{setspace}
\usepackage{booktabs}
\usepackage{array}
\usepackage{cite}
\usepackage{algorithm}
\usepackage{algorithmic}
\usetikzlibrary{patterns}
\usetikzlibrary{graphs}
\usetikzlibrary{graphs.standard}
\usepackage{enumitem}

\newtheorem{theorem}{Theorem}
\newtheorem{definition}{Definition}
\newtheorem{lemma}{Lemma}
\newtheorem{remark}{Remark}
\newtheorem{assumption}{Assumption}

\DeclareMathOperator*{\col}{col}
\DeclareMathOperator*{\adj}{adj}
\DeclareMathOperator*{\R}{\mathcal{R}}
\DeclareMathOperator*{\mS}{\mathcal{S}_\mathcal{R}}

\ifnum\pdfshellescape=1
\fi

\makeatletter

\newcommand{\Rmnum}[1]{\expandafter\@slowromancap\romannumeral #1@}
\makeatother

\title{Resilient Distributed Parameter Estimation in Sensor Networks}

\author{Jiaqi Yan, Kuo Li, and Hideaki Ishii
	\thanks{J. Yan and H. Ishii are with the Department of Computer Science, Tokyo Institute of Technology, Japan. Emails: {jyan@sc.dis.titech.ac.jp, ishii@c.titech.ac.jp}. \; K. Li is with the Center for Intelligent and Networked System, Department of Automation, BNRist, Tsinghua University, P.R. China. Email: li-k19@mails.tsinghua.edu.cn.}%
	\thanks{This work was supported in the part by JSPS under Grants-in-Aid for Scientific Research Grant No. 22H01508 and 21F40376.}
}
\begin{document}
\maketitle

\begin{abstract}
In this paper, we study the problem of parameter estimation in a sensor network, where the measurements and updates of some sensors might be arbitrarily manipulated by adversaries. Despite the presence of such misbehaviors, normally behaving sensors make successive observations of an unknown $d$-dimensional vector parameter and aim to infer its true value by cooperating with their neighbors over a directed communication graph. To this end, by leveraging the so-called dynamic regressor extension and mixing procedure, we transform the problem of estimating the  vector parameter to that of estimating $d$ scalar ones. For each of the scalar problem, 
we propose a resilient combine-then-adapt diffusion algorithm, where each normal sensor performs a resilient combination to discard the suspicious estimates in its neighborhood and to fuse the remaining values, alongside an adaptation step to process its streaming observations. With a low computational cost, this estimator guarantees that each normal sensor exponentially infers the true parameter even if some of them are not sufficiently excited.
\end{abstract}

\section{Introduction}

As a fundamental problem in various applications such as system identification and adaptive control, distributed parameter estimation has been studied in the literature over decades (see, e.g., \cite{goodwin2014adaptive,xie2020convergence,schizas2009distributed,chen2013distributed,yan2022distributed}). In this problem, each sensor observes (partial) information of a system with an unknown (vector) parameter, and attempts to consistently estimate the true parameter by cooperating with others.

As for a single sensor, it is well known that consistent estimation is possible only if its regressor meets certain excitation conditions. Moreover, exponential convergence can be further achieved if a \textit{persistent excitation} (PE) condition is verified, which guarantees that the input signals of a plant are sufficiently rich that all modes of the plant can be excited \cite{goodwin2014adaptive}. However, in a distributed framework, the PE condition may not necessarily hold at every sensor side. Therefore, by properly introducing consensus algorithms, weaker excitation conditions have been proposed, with which sensors collectively satisfy the PE condition and cooperatively fulfill the estimation task (see, for example, \cite{chen2013distributed,xie2018analysis,matveev2021diffusion}).

However, distributed estimation algorithms, despite relaxing the condition on PE and providing better flexibility as compared to their centralized counterparts, are vulnerable to adversarial behaviors in the network \cite{chen2019resilient}. As the scale of the network increases, it becomes especially difficult to secure every sensor and communication channel. Particularly, adversaries could manipulate measurements or transmissions of sensors and disrupt the operation of conventional algorithms. In fact, as reported in \cite{mitra2019byzantine}, even a single adversary is able to drive normal sensors falsely to its desired estimates.

Inspired by these issues, recent research efforts have been devoted to the design of secure estimation protocols. A large class of methods focuses on developing resilient algorithms which ensure that all the normally behaving sensors resiliently recover the unknown parameter even in the presence of attacks. To raise the resiliency against malicious behaviors, many approaches adopt the idea to simply ignore the most extreme values in the neighborhood. Stemming from this idea, a family of strategies termed as mean-subsequence reduced (MSR) algorithms has been proposed and widely applied to solve the problem of resilient consensus (see, for example, \cite{dolev1986reaching,leblanc2013resilient,dibaji2017resilient,ishii2022overview}). However,
different from the consensus problem, which does not incorporate sensors' measurements, in the problem of resilient estimation, each sensor must generate a secure estimate by processing its streaming data. To cope with this issue, the works \cite{leblanc2014resilient,mitra2019byzantine} have developed resilient estimators by extending the MSR algorithms. These estimators are shown to be resilient to Byzantine sensors if the network is sufficiently connected and the collective measurements from normal sensors are observable for the system state. Other resilient approaches include \cite{chen2019resilient,meng2021distributed}, where a pre-defined threshold is necessary at each sensor side to check the distance between the local estimates of its neighbors and its own, and thus limits effects from the misbehaving ones.

In this paper, we also investigate the problem of distributed parameter estimation in an adversarial environment. However, different from most of the aforementioned works, where the regressor is assumed to be a constant matrix, we consider a more general model. Specifically, the observations of each sensor are generated via a linear regression equation (LRE), which is able to describe the input-output relationship of a large class of 
linear and nonlinear dynamical systems \cite{goodwin2014adaptive}. Moreover, notice that in the literature \cite{chen2019resilient,meng2021distributed,chen2018resilient}, it is assumed that only the measurements of sensors can be manipulated. In contrast, this paper considers the case where not only the measurements, but also 
the updating rules of a sensor can be faulty and arbitrarily manipulated. Notice that the latter situation could happen if some ``non-participating" sensors exist in the network, which weigh their private interests more than the public ones and are not willing to follow the given protocols. 

Our estimator is developed inspired by the so-called dynamic regressor extension and mixing (DREM) algorithm. The DREM was first introduced in the recent work \cite{aranovskiy2017performance}, where, in a fault-free environment, it reveals decent performance in relaxing the excitation condition and guaranteeing asymptotic convergence at a fusion center. This paper, 
with subtle modifications on DREM, proposes a \textit{resilient combine-then-adapt} (RCTA) diffusion algorithm to accommodate the difficulties brought by the distributed framework and malicious behaviors. 
To be specific, by leveraging DREM, we transform the problem of estimating a $d$-dimensional vector parameter to that of $d$ scalar ones: one for each of the unknown parameters. For each of the scalar problem, an estimation strategy is given, where each normal sensor runs a resilient algorithm to discard the suspicious estimates in its in-neighborhood and fuse the remaining values, alongside an adaptation process to incorporate its own measurements by using least-mean squares. We provide sufficient conditions under which each normal sensor exponentially infers the true parameter over a directed graph, even if some of them are not sufficiently excited through their inputs. 
Notice that by decoupling the vector estimation problem into scalar ones, the algorithm proposed here reveals a lightweight implementation, which yields lower cost in solving the more complicated distributed estimation problems. 

The rest pf this paper is organized as follows. We will introduce some preliminaries and problem formulation in Sections~\ref{sec:pre} and \ref{sec:form}, respectively. The main algorithm and its convergence analysis are presented in Section~\ref{sec:main}. We test the main results through some numerical examples in Section~\ref{sec:simulation}. Finally, Section~\ref{sec:conclude} concludes the paper.

\section{Preliminaries}\label{sec:pre}
Consider a digraph $\mathcal{G}=(\mathcal{V},\mathcal{E})$. Let $\mathcal{V}$ be the set of sensors, and $\mathcal{E}\subseteq \mathcal{V}\times\mathcal{V}$ be the set of edges. An edge from sensor $j$ to $i$ is denoted by $e_{ij}\in \mathcal{E}$, indicating sensor $i$ can receive the information directly from sensor $j$. Accordingly, the sets of in-neighbors and out-neighbors of agent $i\in \mathcal{V}$ are defined, respectively, as 
\begin{equation}
\mathcal{N}_i^+\triangleq\{j\in \mathcal{V}|e_{ij}\in \mathcal{E}\},\ \mathcal{N}_i^-\triangleq\{j\in \mathcal{V}|e_{ji}\in \mathcal{E}\}.
\end{equation}
 For the algorithms employed in this paper, we shall characterize their resilience in terms of the definitions below \cite{mitra2019byzantine}:

\begin{definition}[$r$-reachable]\label{def:reachable}
	Consider the digraph $\mathcal{G}=(\mathcal{V},\mathcal{E})$. A set $\mathcal{S}\subseteq \mathcal{V}$ is said to be $r$-reachable if it contains at least one sensor that has at least $r$ in-neighbors from outside $\mathcal{S}$. That is, there exists $i\in\mathcal{S}$ such that $|\mathcal{N}_i^+\backslash \mathcal{S}| \geq r$. 
\end{definition}

\begin{definition}[Strongly $r$-robust w.r.t. $\mathcal{S}$]\label{def:robust_wrt_set}
	Consider the digraph $\mathcal{G}=(\mathcal{V},\mathcal{E})$ with $\mathcal{S}\subseteq \mathcal{V}$ being a nonempty subset of $\mathcal{V}$. The graph $\mathcal{G}$ is said to be strongly $r$-robust w.r.t. $\mathcal{S}$, if for any nonempty subset $\mathcal{S}'\subseteq \mathcal{V}\backslash\mathcal{S}$, $\mathcal{S}'$ is $r$-reachable.
\end{definition}

Intuitively, Definition \ref{def:robust_wrt_set} requires that for any nonempty subset of $\mathcal{V}\backslash\mathcal{S}$, it has
at least one sensor with a sufficient number of in-neighbors from outside. 
\section{Problem Formulation}\label{sec:form}

Let us consider the problem of distributed parameter estimation, where multiple sensors cooperatively infer an \textit{unknown} parameter through local measurements. To be specific, for each sensor $i\in\{1,2,...,m\}$, the \textit{measurable} signals $y_i(k)\in\mathbb{R}$ and $\phi_i(k)\in \mathbb{R}^d$ are related via the following linear regression equation (LRE)\footnote{The results in this paper can be readily generalized to cases where the sensor outputs a vector measurement $y_i(k)$, by treating each of its entries independently as a scalar measurement.}:
\begin{equation}\label{eqn:LRE}
y_i(k)= \theta^\prime \phi_i(k),
\end{equation}
where $\theta \in \mathbb{R}^d$ is the parameter to be estimated and $\theta^\prime$ denotes its transpose. 

The sensors aim to estimate $\theta$ from a stream of measurable signals. However, in a practical network, a single sensor may not be sufficiently excited through its inputs. Therefore, the signals available at its local side are not enough to consistently estimate the parameter $\theta$ \cite{goodwin2014adaptive}. In this respect, each sensor intends to obtain an exact estimate on $\theta$ through the information exchange with others. We use the digraph $\mathcal{G}=(\mathcal{V},\mathcal{E})$ to model the interaction among them.

%

\subsection{Attack model}
This paper is concerned with the parameter estimation in an adversarial environment, where some of the sensors might be faulty or misbehaving. Let us denote the set of indices of these sensors as $\mathcal{F}\subset \mathcal{V}$. Any sensor $i\in \mathcal{F}$ could be the one that fails to follow the pre-defined estimation protocol or 
 whose transmitted data is manipulated
by an adversary. On the other hand, the normal or benign sensors will always adopt the prescribed estimation algorithm;
the set of such nodes is denoted by $\mathcal{R}$. Given the limited energy of adversaries, it is reasonable to assume an upper bound on the number of faulty sensors. In this paper, we shall consider an $f$-local attack model as defined below\footnote{Note that the $f$-local attack model assumed here is more general than the $f$-total attack model, where the total number of faulty sensors is assumed to be upper bounded by $f$.}:
\begin{assumption}[Attack model]\label{assm:attack}
The network $\mathcal{G}=(\mathcal{V},\mathcal{E})$ is under an $f$-local attack model. That is, for any normal sensor $i\in\R$, it has no more than $f$ in-neighbors that are misbehaving, i.e., $|\mathcal{F}\cap\mathcal{N}^+_i|\leq f$.
\end{assumption}

\subsection{Resilient parameter estimation}
 Our goal is to ensure that
all of the normal sensors consistently estimate the parameter $\theta$. Specifically, let $\hat{\theta}_i(k)\in\mathbb{R}^d$ be the local estimate produced by sensor $i$ at time $k$. This paper aims to develop a distributed estimation algorithm which works resiliently against the set of misbehaving sensors by solving the following problem:

\begin{definition}[Resilient parameter estimation]
Resilient parameter estimation is said to be achieved if the local estimate of each normal sensor converges to the true parameter $\theta$, regardless of the initial states and network misbehaviors, namely, $$\displaystyle \lim_{k\to\infty}  \hat{\theta}_i(k) =\theta, \;\forall i\in\R.$$
\end{definition}


\vspace{10pt}
\section{Main Results}\label{sec:main}
This section will provide a resilient algorithm to solve the problem of distributed parameter estimation. Specifically, by decoupling the $d$-dimensional estimation problem into $d$ scalar ones, every normal sensor performs $d$ independent MSR-based estimation algorithms in a parallel manner, each of which infers an entry of the unknown vector parameter. We shall show that this protocol is lightweight and efficient even in the presence of faulty sensors.

\subsection{Dynamic regressor extension and mixing (DREM)}
Our estimator is developed based on the \textit{dynamic regressor extension and mixing} (DREM) algorithm, which was first introduced in \cite{aranovskiy2017performance}.
The DREM algorithm is expressed by the three new variables for each sensor $i\in\mathcal{V}$, defined as follows:
\begin{equation}\label{eqn:definition}  
\begin{split}
\Phi_{i}(k)&\triangleq\begin{bmatrix}
	\left(\phi_{i}(k)\right)^\prime \\
	\left(\phi_{i}(k-1)\right)^\prime \\
	\vdots \\
	\left(\phi_{i}(k-d+1)\right)^\prime
\end{bmatrix}\in\mathbb{R}^{d\times d}, \\\overline{y}_{i}(k) &\triangleq\adj(\Phi_{i}(t))\left[\begin{array}{c}
	y_{i}(k) \\
	y_{i}(k-1) \\
	\vdots \\
	y_{i}(k-d+1)
\end{array}\right]\in\mathbb{R}^{d},\\
\delta_{i}(k)&\triangleq \det(\Phi_{i}(k)),
\end{split}                      
\end{equation}
where we respectively denote by $\adj(\Phi_{i}(k))$ and $\det(\Phi_{i}(k))$ the adjugate matrix and determinant of matrix $\Phi_{i}(k)$. Notice that $\overline{y}_{i}(k)$ is a $d$-dimensional column vector. For simplicity, we denote by $\overline{y}^\ell_{i}(k)$ the $\ell$-th entry of it. By using DREM, the following lemma is obtained:

\begin{lemma}[\hspace{1pt}\cite{aranovskiy2017performance}]\label{lmm:Y}
	Consider the LRE \eqref{eqn:LRE}. For any $\ell\in\{1,\cdots,d\},$ it holds for any $i\in\mathcal{V}$ and at any time $k$ that
	\begin{equation}\label{eqn:scalarLRE}
		\overline{y}_i^\ell(k) = \delta_{i}(k)\theta^\ell,
	\end{equation}
where $\overline{y}_i(k)$ and $\delta_{i}(k)$ are given in \eqref{eqn:definition}. Moreover, $\theta^\ell$ is the $\ell$-th entry of the true parameter $\theta$.
\end{lemma}

Therefore, leveraging DREM, we generate $d$ scalar LREs as presented in \eqref{eqn:scalarLRE}: one for each of the unknown parameters. 

%

\subsection{Description of the resilient algorithm}
This subsection is devoted to developing the resilient estimator based on DREM. In order to reject the possible attacks, we propose a \textit{resilient combine-then-adapt} (RCTA) diffusion algorithm, which overrules the malicious effects from faulty sensors by performing a resilient convex combination and then an adaptation. 
Specifically, each normal sensor $i\in \mathcal{R}$ starts with any initial estimate $\hat\theta_i(0)\in \mathbb{R}^d$. At any time $k>0$, it makes an estimation as in Algorithm \ref{alg:resilient}. 

\begin{algorithm}[h!] 
1:\: Collect the local estimates from all in-neighboring agents $j\in\mathcal{N}_i^+$, and place these values in the multiset $\mathcal{X}_i(k)$.

2:\: \textbf{for} $\ell\in\{1,2,...,d\}$ \textbf{do}
\qquad \begin{enumerate}[leftmargin = 30 pt]
	\item[a):] Set $\mathcal{J}^{\ell}_i(k)=\mathcal{X}_i(k)$. Then sort the points in $ \mathcal{J}^{\ell}_i(k)$ according to their $\ell$-th entries in an ascending order.
	\item[b):] Based on the sorted set, remove the first $f$ points from $\mathcal{J}^{\ell}_i(k)$ which have the smallest $\ell$-th entries. If there are less than $f$ points in $\mathcal{J}^{\ell}_i(k)$, then remove all of them.
	\item[c):] Similarly, remove the last $f$ points from $\mathcal{J}^{\ell}_i(k)$ which have the largest $\ell$-th entries. If there are less than $f$ points in $\mathcal{J}^{\ell}_i(k)$, then remove all of them.
	\item[d):] Sensor $i$ resiliently combines the neighboring estimates as
	
	\textit{(Resilient combination)  } 
	\begin{equation}\label{eqn:resilientcomb}
		\begin{split}
		&\bar{\theta}_i^\ell (k) \\&=\begin{cases}
			\hat{\theta}_i^\ell (k), \quad \text{if } \mathcal{J}^{\ell}_i(k)=\varnothing,\\
		a^\ell_{ii}(k)\hat{\theta}_i^\ell (k)+\sum_{j\in\mathcal{J}^{\ell}_i(k)}a^\ell_{ij}(k)\hat{\theta}_j^\ell (k), \;\; \text{otherwise},
		\end{cases}
		\end{split}
	\end{equation}
	where each weight $a^\ell_{ij}(k)$ is lower bounded by $\alpha>0$ and $a^\ell_{ii}(k)+\sum_{j\in\mathcal{J}^{\ell}_i(k)}a^\ell_{ij}(k)=1$. 
	\item[e):] Sensor $i$ updates the $\ell$-th entry of local estimate as
	
	\textit{(Adaptation)  }
	\begin{equation}\label{eqn:update}
	\begin{split}
	&\hat{\theta}^{\ell}_{i}(k+1)\\&\;=\bar{\theta}_i^\ell (k)
	+\frac{\delta_{i}(k)}{\mu_{i}+\left(\delta_{i}(k)\right)^{2}}\left(\overline{y}^{\ell}_{i}(k)-\delta_{i}(k) \bar{\theta}_i^\ell (k)\right),
	\end{split}
	\end{equation}
	where $\mu_{i}>0$.
\end{enumerate}
\quad \textbf{end for}

3:\: Transmit $\hat{\theta}_{i}(k+1)=\col(\hat{\theta}^{\ell}_{i}(k+1))$ to out-neighbors. \\
	\caption{Resilient parameter estimation algorithm}
	\label{alg:resilient}
\end{algorithm}

In the proposed algorithm, for inferring each entry of the unknown parameter, i.e., $\theta^\ell$, each normal sensor $i$ sorts the received estimates based on their $\ell$-th entries, where it discards up to $f$ smallest ones and up to $f$ largest ones. As discussed previously, for normal sensor $i\in \mathcal{R}\backslash\mS$, it has at least $3f+1$ in-neighbors. Since it removes at most $2f$ values at each $\ell\in\{1,\cdots,d\}$, $\mathcal{J}^{\ell}_i(k)$ must be non-empty at any time $k$. On the other hand, if $i\in\mS$, it is possible that $\mathcal{J}^{\ell}_i(k)=\varnothing$. In either case, sensor $i$ calculates a resilient combination \eqref{eqn:resilientcomb} via linearly combining its own estimate and the ones in $\mathcal{J}^{\ell}_i(k)$. After that, an adaptation \eqref{eqn:update} is performed using a least-mean square scheme to update sensor $i$'s local estimate on $\theta^\ell$. Finally, sensor $i$ aggregates its estimates on all the entries and sends them to its out-neighbors. 


In contrast, a main feature of Algorithm~\ref{alg:resilient} is its simplicity in implementation and computation. Particularly, by leveraging DREM, we decouple the $d$-dimensional parameter estimation problem into $d$ scalar ones. Then the resilient combination \eqref{eqn:resilientcomb} is performed within each scalar system, where only coordinate-wise message sorting and trimming are needed. The computational complexity of each normal sensor $i$ can be shown to be $\mathcal{O}(d\cdot n_i\log_2n_i)$, where $n_i\triangleq|\mathcal{N}_i^+|$. As compared with the existing solutions in multi-dimensional spaces, such as \cite{wang2018resilient,vaidya2012iterative,mendes2013multidimensional,yan2020resilient}, Algorithm~\ref{alg:resilient} yields lower computational cost. 

\subsection{Assumptions} 
Before proceeding, we first introduce assumptions that would be adopted in this paper.

It is well known that consistent estimation is possible only if the input signal satisfies certain excitation conditions. In particular, a 
\textit{persistent excitation} (PE) condition, which guarantees the signals to be sufficiently rich, is usually required to achieve an exponential convergence (\hspace{1pt}\cite{goodwin2014adaptive}). 
On the other hand, in order to countermeasure the faulty behaviors in an adversarial environment, the network should contain a certain degree of redundancy in its communication structure. In this respect, let us introduce the following assumptions:
\begin{assumption} \label{assm:topology} 
	There exists a subset of sensors $\mathcal{S}\subset\mathcal{V}$ such that the following statements hold:
	\begin{enumerate}
		\item {[Persistent excitation]} There exist $\Delta>0$ and a finite time $T\in\mathbb{N}_{+}$ such that the following PE condition is satisfied by each $i\in\mathcal{S}$:
		\begin{equation}\label{eqn:PE}
		\sum_{t=k}^{k+T-1}(\delta_{i}(t))^2\geq \Delta, \;\forall k.
		\end{equation} 
		\item {[Network topology]} The network $\mathcal{G}=(\mathcal{V},\mathcal{E})$ is strongly $(3f+1)$-robust w.r.t. $\mathcal{S}$. 
	\end{enumerate}
\end{assumption}

According to Assumption~\ref{assm:topology}, the set $\mathcal{S}$ consists of the sensors that are persistently excited. Moreover, there could exist two cases: all sensors in the network are persistently excited and thus $\mathcal{S}=\mathcal{V}$, or only a subset of the sensors is persistently excited, but they are ``sufficiently connected" to all the others. Particularly, as for the second case, since the network is strongly $(3f+1)$-robust w.r.t. $\mathcal{S}$, one can verify from Definition~\ref{def:robust_wrt_set} that $|\mathcal{S}|\geq 3f+1$. Moreover, each sensor $i\in \mathcal{V}\backslash\mathcal{S}$ has at least $3f+1$ in-neighbors\footnote{To see this, suppose that sensor $i$ outside the set $\mathcal{S}$ has at most $3f$ in-neighbors. Let us choose $\mathcal{S}' = \{i\}$. Then based on Definition \ref{def:robust_wrt_set}, $\mathcal{S}'$ is at most $3f$-reachable. Therefore, the network can not be strongly $(3f+1)$-robust w.r.t. $\mathcal{S}$, which contradicts the assumption.}. As will be proved later, in either case, our algorithm guarantees that each normal sensor, whether it satisfies the PE condition or not, consistently estimates the true parameter.

Notice that sensors in $\mathcal{S}$ may also be faulty or misbehaving. For simplicity, we denote by $\mS$ the subset of $\mathcal{S}$ which contains only the normal sensors, that is, $\mS\triangleq \mathcal{S}\cap \mathcal{R}$. The following assumption is finally made:

\begin{assumption}\label{assm:SR}
	The set $\mS$ is nonempty.
\end{assumption}

Assumption~\ref{assm:SR} implies that at least one normal sensor exists, which can find the true parameter. Notice that this is necessary for $\theta$ to be consistently estimated by all normal sensors. 

\subsection{Convergence analysis}
To theoretically analyze the performance of Algorithm~\ref{alg:resilient}, let us define the following set for each normal sensor $i\in\mathcal{R}$:
\begin{equation}
	\widetilde{\mathcal{R}}_{i} \triangleq (\mathcal{N}_i^+\cap \mathcal{R})\cup\{i\},
\end{equation}
which includes all the normal in-neighbors of sensor $i$ and itself. This set is used in our analysis only and need not be known by sensor $i$. In the next lemma, we shall prove that $\bar{\theta}_i^\ell (k)$ obtained by performing the resilient combination \eqref{eqn:resilientcomb} is indeed a convex combination of the local estimates from the normal sensors in $\widetilde{\mathcal{R}}_{i}$: 

\begin{lemma}\label{lmm:equalform}
Suppose that under Assumptions~\ref{assm:attack}--\ref{assm:SR}, each normal sensor performs parameter estimation by following Algorithm~\ref{alg:resilient}. Then, for any normal sensor $i\in\mathcal{R}$ and $\ell\in\{1,2,...,d\}$, there exists a set of weights $\{d^\ell_{ij}(k)\}$ such that $\bar{\theta}_i^\ell (k)$ can be represented as 
\begin{equation}\label{eqn:equalform}
	\bar{\theta}_i^\ell (k) =  \sum_{j\in\widetilde{\mathcal{R}}_{i}}d^\ell_{ij}(k)\hat{\theta}_j^\ell (k),
\end{equation}
where $\bar{\theta}_i^\ell (k)$ is calculated by \eqref{eqn:resilientcomb}. Moreover, the following statements hold:
\begin{enumerate}
	\item $\bar{\theta}_i^\ell (k)$ is a convex combination of its own estimate and the ones received from normal in-neighbors. That is, each weight in \eqref{eqn:equalform} is non-negative and $\sum_{j\in\widetilde{\mathcal{R}}_{i}}d^\ell_{ij}(k)=1$;
	\item $d^\ell_{ii}(k) \geq \alpha$;  
	\item For any normal sensor $j$ whose estimate at $\ell$-th entry is retained by sensor $i$, i.e., $j\in \mathcal{J}^{\ell}_i(k)\cap\mathcal{R}$, it follows that $ j\in\widetilde{\mathcal{R}}_{i}$ and $d^\ell_{ij}(k)\geq \alpha$. 
	
\end{enumerate}
\end{lemma}
%

As implied by Lemma~\ref{lmm:equalform}, $\bar{\theta}_i^\ell(k)$ is updated in a safe manner as it always remains in a region that is determined by the local estimates of only normal sensors. Hence, by resiliently combining in-neighbors' estimates, \eqref{eqn:resilientcomb} prevents the faulty sensors from taking arbitrary control over the dynamics of any normal one.

Therefore, one concludes that the resilient combination step ensures that each normal sensor is updated in a safe manner. We shall next study the performance of Algorithm~\ref{alg:resilient} on achieving the parameter estimation. This is particularly guaranteed by the adaptation step in \eqref{eqn:update}.

To see this, for each sensor $i\in\mathcal{V}$, let us define the estimation error of it at $\ell$-th entry as
\begin{equation}
\tilde{\theta}_i ^\ell(k) \triangleq \hat{\theta}_i ^\ell(k)-\theta^\ell.
\end{equation}
Combining \eqref{eqn:scalarLRE} and \eqref{eqn:update} with Lemmas~\ref{lmm:Y} and \ref{lmm:equalform}, the dynamics of $\tilde{\theta}_i^\ell(k)$ is obtained as 
\begin{equation}\label{eqn:error}
\begin{split}
&\tilde{\theta}_i ^\ell(k+1)=\hat{\theta}^{\ell}_{i}(k+1)-\theta^\ell\\&=\sum_{j\in\widetilde{\mathcal{R}}_{i}}d^\ell_{ij}(k)\hat{\theta}_j^\ell (k)-\theta^\ell
\\&\qquad+\frac{(\delta_{i}(k))^2}{\mu_{i}+\left(\delta_{i}(k)\right)^{2}}\left(\theta^\ell- \sum_{j\in\widetilde{\mathcal{R}}_{i}}d^\ell_{ij}(k)\hat{\theta}_j^\ell (k)\right)\\&=
\left(1-\frac{(\delta_{i}(k))^2}{\mu_{i}+\left(\delta^\ell_{i}(k)\right)^{2}}\right)\sum_{j\in\widetilde{\mathcal{R}}_{i}} d^\ell_{ij}(k)(\hat{\theta}_j^\ell (k)-\theta^\ell)\\&=\nu_i(k)\sum_{j\in\widetilde{\mathcal{R}}_{i}} d^\ell_{ij}(k)\tilde{\theta}_j ^\ell(k),
\end{split}
\end{equation}
where
$
	\nu_i(k)\triangleq1-\frac{(\delta_{i}(k))^2}{\mu_{i}+\left(\delta_{i}(k)\right)^{2}}\leq 1.
$


Then let us introduce the following results, the proofs of which are omitted due to the space limitation:
\begin{lemma}
Consider the network of sensors verifying the LRE \eqref{eqn:LRE}. Suppose under Assumptions~\ref{assm:attack}--\ref{assm:SR}, that each normal sensor performs parameter estimation by following Algorithm~\ref{alg:resilient}.
Then it holds for any $\ell$ that
\begin{equation}\label{eqn:m}
	m(k+1)\leq m(k), \;\forall k.
\end{equation}
\end{lemma}
%
Therefore, the maximum estimation error would never increase throughout the execution. Further, 
we show that this error will definitely decrease after a finite time:

\begin{lemma}\label{lmm:m}
Consider the network of sensors verifying the LRE \eqref{eqn:LRE}. Suppose under Assumptions~\ref{assm:attack}--\ref{assm:SR}, that each normal sensor performs parameter estimation by following Algorithm~\ref{alg:resilient}. Then, it holds for any $\ell$ that
	\begin{equation}\label{eqn:t+T+M}
		m(k+T+M+1)\leq (1-\alpha^{T+M}(1-\gamma))m(k).
	\end{equation}
\end{lemma}

\medskip
With the above preparations, we are now ready to state the main result of the paper on the convergence of the estimation error as follows:

\begin{theorem}\label{thm:converge}
Consider the network of sensors satisfying the LRE \eqref{eqn:LRE}.
Suppose that Assumptions~\ref{assm:attack}--\ref{assm:SR} hold. Let each normal sensor update with Algorithm~\ref{alg:resilient}. Then resilient parameter estimation is achieved with an exponential rate. Namely, it holds that
\begin{equation}
\lim\limits_{k\to \infty} \tilde{\theta}_i(k) = 0,\;\forall i\in\R,
\end{equation}
regardless of the network misbehaviors.
\end{theorem}

In view of Theorem~\ref{thm:converge}, the convergence of Algorithm~\ref{alg:resilient} holds regardless of the presence and behaviors of faulty sensors. Therefore, the algorithm can be applied to withstand even the worst-case adversaries.

\begin{remark}
We further point out that the topology condition that the network $\mathcal{G}=(\mathcal{V},\mathcal{E})$ is strongly $(3f+1)$-robust w.r.t. a set $\mathcal{S}$ is similar to those in the works \cite{usevitch2019resilient,mitra2019byzantine,yan2021resilient}. In these works, agents in $\mathcal{S}$ are assumed to have direct knowledge of the state to be estimated. Thus, these agents perform as leaders while others are followers that resiliently track the states of the leaders. Note that in the framework of leader-following consensus, the leaders only broadcast their estimates to the followers while not taking account of others' states into the updates of their own. 
However, different from such a
framework, in the parameter estimation here, the sensors in $\mathcal{S}$
are the ones that satisfy the PE condition. Moreover, according
to \eqref{eqn:resilientcomb}, the dynamics of them are inevitably affected by other
sensors including the ones outside $\mathcal{S}$. Therefore, the existing methods for convergence analysis are prevented from being directly applied. In order to solve this problem, the adaptation step \eqref{eqn:update} is necessary to guarantee the convergence of normal sensors.


\end{remark}

\section{Numerical Example}\label{sec:simulation}
\begin{figure}[!htbp]
	\centering
	\includegraphics[width=0.3\textwidth]{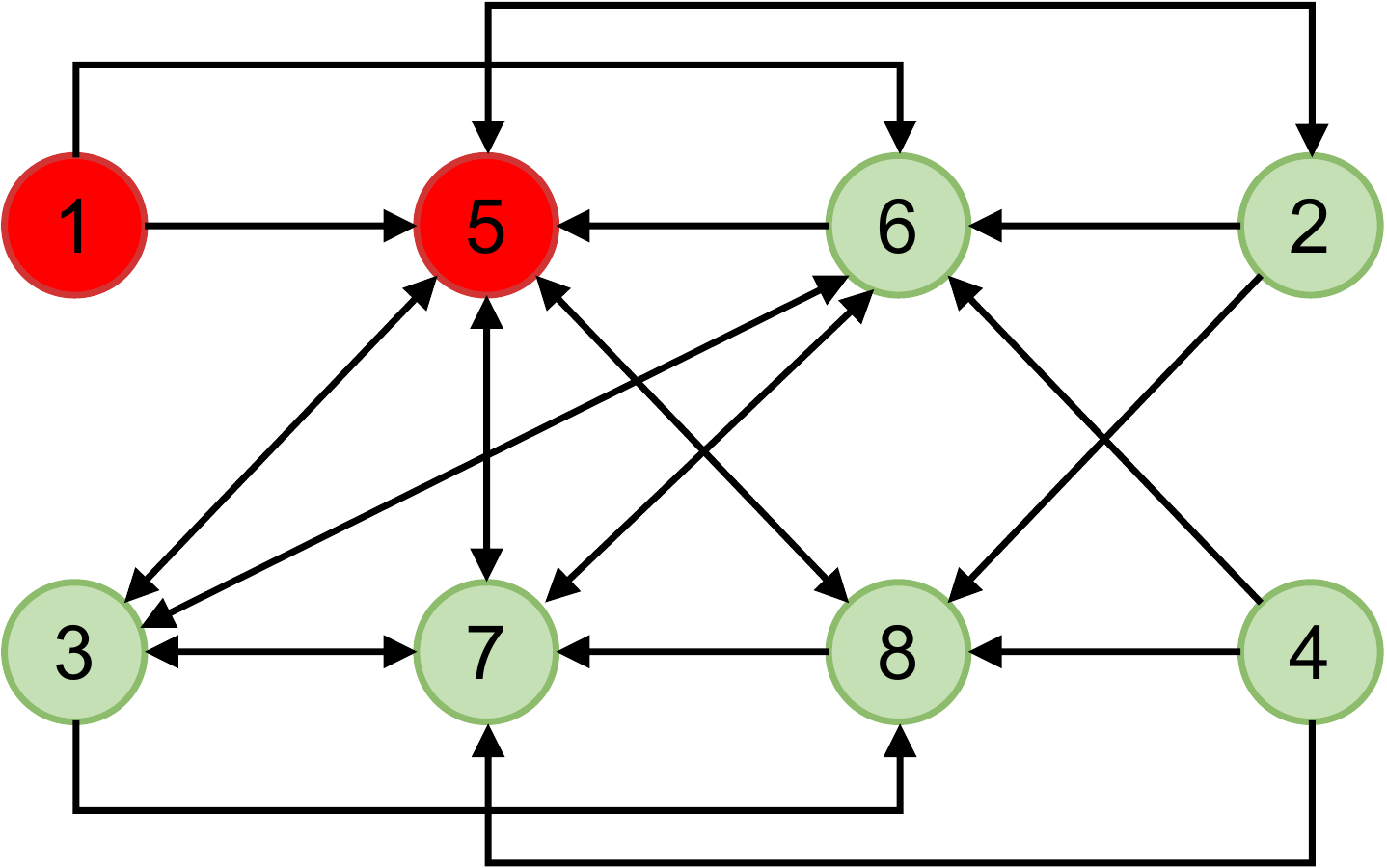}
	\caption{A network that is strongly $4$-robust w.r.t. the set $\mathcal{S}=\{1,2,3,4\}$.}
	\label{fig:network}
\end{figure}

This example considers the network of $8$ sensors, which aim to cooperatively estimate a $2$-dimensional parameter $\theta$ over the digraph given by Fig.~\ref{fig:network}. It can be checked that the graph is strongly $4$-robust w.r.t. the set $\mathcal{S}=\{1,2,3,4\}$. Therefore, according to Theorem~\ref{thm:converge}, it is able to tolerate the $1$-local attack.  

To verify this, let sensors $1$ and $5$ be misbehaving. They intend to prevent the estimation task from being fulfilled by maliciously broadcasting their estimates as 
\begin{equation}
\begin{split}
\hat{\theta}^{1}_{1}(k)&=2,\; \hat{\theta}^{2}_{1}(k)=-2,\\
\hat{\theta}^{1}_{5}(k)&=k/20+2,\; \hat{\theta}^{2}_{5}(k)=0.5\sin(k/5).
\end{split}
\end{equation}
On the other hand, the regressor $\phi_i(t)$ for each normal sensor is given by
\begin{equation}
\begin{split}
\phi_2(t)&=[
	a(t) \quad 1
]^\prime,\;\phi_3(t)=[
1\quad b(t)
]^\prime,\\\phi_4(t)&=
\begin{cases}
	[1 \quad 2 ]^\prime, \text{ if } t \text{ is odd},\\
[2 \quad 3 ]^\prime, \text{ if } t \text{ is even},
\end{cases}
\\\phi_6(t)&=[1 \quad 0 ]^\prime,\;\phi_7(t)=[0 \quad 2 ]^\prime,\;\phi_8(t)=[1 \quad 1 ]^\prime,
\end{split}
\end{equation}
where 
$
	a(t)=a(t-1)+\cos\left(\frac{t\pi}{4}\right), \;b(t)=b(t-1)+\cos\left(\frac{t\pi}{2}\right),
$
with $a(0)=1$ and $b(0)=2$. Notice that the PE condition is satisfied by $\delta_i(k), i=\{2,3,4\}$. We set the initial estimates of all sensors as $[0\quad 0]^\prime$ and other parameters as
$
	\theta = [2.5 \quad  -1]^\prime,$ and $\mu_i = 0.1i, \;\forall i.
$

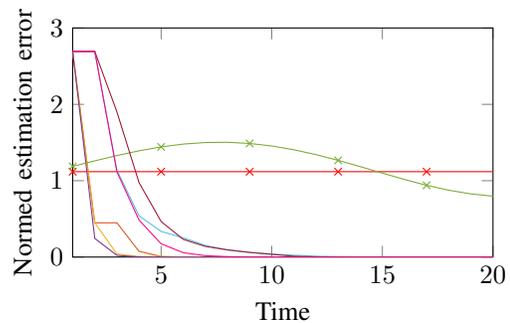
\begin{figure}[!htbp]
	\centering
%
%
\definecolor{mycolor1}{rgb}{1.00000,0.00000,0.50000}%
\definecolor{mycolor2}{rgb}{0.85000,0.32500,0.09800}%
\definecolor{mycolor3}{rgb}{0.92900,0.69400,0.12500}%
\definecolor{mycolor4}{rgb}{0.49400,0.18400,0.55600}%
\definecolor{mycolor5}{rgb}{0.46600,0.67400,0.18800}%
\definecolor{mycolor6}{rgb}{0.30100,0.74500,0.93300}%
\definecolor{mycolor7}{rgb}{0.63500,0.07800,0.18400}%
\begin{tikzpicture}

\begin{axis}[%
width=2.2in,
height=1.2in,
at={(0.758in,0.481in)},
scale only axis,
xmin=1,
xmax=20,
xlabel={{Time}},
ymin=0,
ymax=3,
ylabel={{Normed estimation error}},
axis background/.style={fill=white}
]
\addplot [color=red, forget plot, color=red, mark=x, mark repeat=4]
  table[row sep=crcr]{%
1	1.11803398874989\\
2	1.11803398874989\\
3	1.11803398874989\\
4	1.11803398874989\\
5	1.11803398874989\\
6	1.11803398874989\\
7	1.11803398874989\\
8	1.11803398874989\\
9	1.11803398874989\\
10	1.11803398874989\\
11	1.11803398874989\\
12	1.11803398874989\\
13	1.11803398874989\\
14	1.11803398874989\\
15	1.11803398874989\\
16	1.11803398874989\\
17	1.11803398874989\\
18	1.11803398874989\\
19	1.11803398874989\\
20	1.11803398874989\\
21	1.11803398874989\\
22	1.11803398874989\\
23	1.11803398874989\\
24	1.11803398874989\\
25	1.11803398874989\\
26	1.11803398874989\\
27	1.11803398874989\\
28	1.11803398874989\\
29	1.11803398874989\\
30	1.11803398874989\\
31	1.11803398874989\\
32	1.11803398874989\\
33	1.11803398874989\\
34	1.11803398874989\\
35	1.11803398874989\\
36	1.11803398874989\\
37	1.11803398874989\\
38	1.11803398874989\\
39	1.11803398874989\\
40	1.11803398874989\\
41	1.11803398874989\\
42	1.11803398874989\\
43	1.11803398874989\\
44	1.11803398874989\\
45	1.11803398874989\\
46	1.11803398874989\\
47	1.11803398874989\\
48	1.11803398874989\\
49	1.11803398874989\\
50	1.11803398874989\\
};
\addplot [color=mycolor2, forget plot]
  table[row sep=crcr]{%
1	2.69258240356725\\
2	0.448763733927876\\
3	0.448763733927876\\
4	0.0747939556546457\\
5	0.00679945051405887\\
6	0.00113324175234315\\
7	0.00113324175234315\\
8	0.000188873625390374\\
9	1.71703295808119e-05\\
10	2.86172159691194e-06\\
11	2.86172159691194e-06\\
12	4.76953599691486e-07\\
13	4.33594182549789e-08\\
14	7.22656987409374e-09\\
15	7.22656987409374e-09\\
16	1.20442836732582e-09\\
17	1.0949354416508e-10\\
18	1.82489927486348e-11\\
19	1.82489927486348e-11\\
20	3.04148504749486e-12\\
21	2.76464905621724e-13\\
22	4.59331883807359e-14\\
23	4.59331883807359e-14\\
24	7.66938336801217e-15\\
25	4.96506830649455e-16\\
26	1.11022302462516e-16\\
27	1.11022302462516e-16\\
28	0\\
29	0\\
30	0\\
31	0\\
32	0\\
33	0\\
34	0\\
35	0\\
36	0\\
37	0\\
38	0\\
39	0\\
40	0\\
41	0\\
42	0\\
43	0\\
44	0\\
45	0\\
46	0\\
47	0\\
48	0\\
49	0\\
50	0\\
};
\addplot [color=mycolor3, forget plot]
  table[row sep=crcr]{%
1	2.69258240356725\\
2	0.448763733927875\\
3	0.0407967030843525\\
4	0.00679945051405887\\
5	0.00679945051405887\\
6	0.00113324175234315\\
7	0.00010302197748582\\
8	1.71703295808531e-05\\
9	1.71703295808531e-05\\
10	2.86172159695317e-06\\
11	2.60156508787686e-07\\
12	4.33594182962115e-08\\
13	4.33594182962115e-08\\
14	7.22656987409374e-09\\
15	6.56960976361759e-10\\
16	1.0949354416508e-10\\
17	1.0949354416508e-10\\
18	1.82489927486348e-11\\
19	1.65920175611054e-12\\
20	2.76506165169824e-13\\
21	2.76506165169824e-13\\
22	4.59331883807359e-14\\
23	4.28839362150501e-15\\
24	4.96506830649455e-16\\
25	4.96506830649455e-16\\
26	1.11022302462516e-16\\
27	0\\
28	0\\
29	0\\
30	0\\
31	0\\
32	0\\
33	0\\
34	0\\
35	0\\
36	0\\
37	0\\
38	0\\
39	0\\
40	0\\
41	0\\
42	0\\
43	0\\
44	0\\
45	0\\
46	0\\
47	0\\
48	0\\
49	0\\
50	0\\
};
\addplot [color=mycolor4, forget plot]
  table[row sep=crcr]{%
1	2.69258240356725\\
2	0.244780218506114\\
3	0.0222527471369197\\
4	0.0020229770124473\\
5	0.000183907001131599\\
6	1.67188182844996e-05\\
7	1.51989257120195e-06\\
8	1.38172052039902e-07\\
9	1.25610957037143e-08\\
10	1.14191765255512e-09\\
11	1.03810901850071e-10\\
12	9.43732847477492e-12\\
13	8.58051405911139e-13\\
14	7.79709400972601e-14\\
15	7.17448075233818e-15\\
16	4.96506830649455e-16\\
17	0\\
18	0\\
19	0\\
20	0\\
21	0\\
22	0\\
23	0\\
24	0\\
25	0\\
26	0\\
27	0\\
28	0\\
29	0\\
30	0\\
31	0\\
32	0\\
33	0\\
34	0\\
35	0\\
36	0\\
37	0\\
38	0\\
39	0\\
40	0\\
41	0\\
42	0\\
43	0\\
44	0\\
45	0\\
46	0\\
47	0\\
48	0\\
49	0\\
50	0\\
};
\addplot [color=mycolor5, color =mycolor5, mark=x, mark repeat=4]
  table[row sep=crcr]{%
1	1.18787066069699\\
2	1.25989285403174\\
3	1.32922825507339\\
4	1.39140433777791\\
5	1.44256346112616\\
6	1.47959903366077\\
7	1.50024248792388\\
8	1.50311691162592\\
9	1.48776423969659\\
10	1.45464871341284\\
11	1.40513807597933\\
12	1.34146405379392\\
13	1.26666364434641\\
14	1.18450091403178\\
15	1.0993628710433\\
16	1.01610911900624\\
17	0.939832012478528\\
18	0.87546310167169\\
19	0.827184760941895\\
20	0.79773743106247\\
21	0.787931030833916\\
22	0.796733677503462\\
23	0.821988107603788\\
24	0.861348578183319\\
25	0.912940122062459\\
26	0.975534918113263\\
27	1.04834476712145\\
28	1.1306448399698\\
29	1.22141787151869\\
30	1.31912954517724\\
31	1.42166682432898\\
32	1.52641578023103\\
33	1.63043119106828\\
34	1.73065026856438\\
35	1.82411470210697\\
36	1.90817845951993\\
37	1.98068855607866\\
38	2.04013156313701\\
39	2.08574100705782\\
40	2.11756125806641\\
41	2.13646297691945\\
42	2.14410441245633\\
43	2.14283267480585\\
44	2.13552083191124\\
45	2.12534206581229\\
46	2.11549282498243\\
47	2.10889280924374\\
48	2.10790670463375\\
49	2.1141422867897\\
50	2.12837229623578\\
};
\addplot [color=mycolor6, forget plot]
  table[row sep=crcr]{%
1	2.69258240356725\\
2	2.69258240356725\\
3	1.13969392055622\\
4	0.539795001770185\\
5	0.336108382627311\\
6	0.252544152135376\\
7	0.152574590912992\\
8	0.0933858343605985\\
9	0.0606836253081021\\
10	0.0394489306762493\\
11	0.024334123311678\\
12	0.0111127353262144\\
13	0.00449652723543843\\
14	0.00170348074612224\\
15	0.000619111046518082\\
16	0.000218668172636034\\
17	7.56357588706479e-05\\
18	2.5747834178621e-05\\
19	8.6554286637955e-06\\
20	2.88014059941698e-06\\
21	9.503533489164e-07\\
22	3.11376776346651e-07\\
23	1.01407471009055e-07\\
24	3.28545462612734e-08\\
25	1.05962504902193e-08\\
26	3.40389993814679e-09\\
27	1.08959511164013e-09\\
28	3.47681447880207e-10\\
29	1.10627953700937e-10\\
30	3.51101377389376e-11\\
31	1.11173295011295e-11\\
32	3.5127457130746e-12\\
33	1.10800260082486e-12\\
34	3.49054136598367e-13\\
35	1.09690091018327e-13\\
36	3.46391362882045e-14\\
37	1.11027853438868e-14\\
38	3.55444797896667e-15\\
39	8.95090418262362e-16\\
40	4.57756679852224e-16\\
41	1.11022302462516e-16\\
42	1.11022302462516e-16\\
43	1.11022302462516e-16\\
44	1.11022302462516e-16\\
45	1.11022302462516e-16\\
46	1.11022302462516e-16\\
47	1.11022302462516e-16\\
48	1.11022302462516e-16\\
49	1.11022302462516e-16\\
50	1.11022302462516e-16\\
};
\addplot [color=mycolor7, forget plot]
  table[row sep=crcr]{%
1	2.69258240356725\\
2	2.69258240356725\\
3	1.90183098910264\\
4	0.975601416200584\\
5	0.464694904666855\\
6	0.231957209106188\\
7	0.139472788464699\\
8	0.0940620893681383\\
9	0.0608339354667262\\
10	0.0357431351978783\\
11	0.0109057455368844\\
12	0.00332701959873067\\
13	0.00101476719716769\\
14	0.000309445663674389\\
15	9.43465472370302e-05\\
16	2.87595541416354e-05\\
17	8.76489957205131e-06\\
18	2.67067189000393e-06\\
19	8.13596563475213e-07\\
20	2.47805753203688e-07\\
21	7.54617923373623e-08\\
22	2.29751128103969e-08\\
23	6.99368549727205e-09\\
24	2.12849855458085e-09\\
25	6.47680457789288e-10\\
26	1.97047128309282e-10\\
27	5.99379031858305e-11\\
28	1.82288202346224e-11\\
29	5.54327196289018e-12\\
30	1.68562050791858e-12\\
31	5.12511705433744e-13\\
32	1.55819633410295e-13\\
33	4.76403413939009e-14\\
34	1.44192285623161e-14\\
35	4.50838044768686e-15\\
36	1.35064460289285e-15\\
37	4.57756679852224e-16\\
38	1.11022302462516e-16\\
39	1.11022302462516e-16\\
40	1.11022302462516e-16\\
41	1.11022302462516e-16\\
42	1.11022302462516e-16\\
43	1.11022302462516e-16\\
44	1.11022302462516e-16\\
45	1.11022302462516e-16\\
46	1.11022302462516e-16\\
47	1.11022302462516e-16\\
48	1.11022302462516e-16\\
49	1.11022302462516e-16\\
50	1.11022302462516e-16\\
};
\addplot [color=mycolor1, forget plot]
  table[row sep=crcr]{%
1	2.69258240356725\\
2	2.69258240356725\\
3	1.11649532267518\\
4	0.484719189732424\\
5	0.173969040290087\\
6	0.056949764587247\\
7	0.0178781164470164\\
8	0.00579608761231149\\
9	0.00181093579978104\\
10	0.000555299051886094\\
11	0.000167734278834524\\
12	5.05555618095562e-05\\
13	1.52061770970626e-05\\
14	4.57924890715299e-06\\
15	1.37667459908439e-06\\
16	4.13604837073768e-07\\
17	1.24182090094074e-07\\
18	3.72986094179472e-08\\
19	1.11969135572821e-08\\
20	3.36060208556998e-09\\
21	1.00843545115285e-09\\
22	3.02641853092917e-10\\
23	9.0811435576202e-11\\
24	2.72474273114913e-11\\
25	8.17464470839168e-12\\
26	2.45242386762323e-12\\
27	7.35798330765779e-13\\
28	2.21037011850425e-13\\
29	6.64191794380373e-14\\
30	1.98850830710367e-14\\
31	5.86214578321713e-15\\
32	1.93892115658267e-15\\
33	4.96506830649455e-16\\
34	1.11022302462516e-16\\
35	1.11022302462516e-16\\
36	1.11022302462516e-16\\
37	1.11022302462516e-16\\
38	1.11022302462516e-16\\
39	1.11022302462516e-16\\
40	1.11022302462516e-16\\
41	1.11022302462516e-16\\
42	1.11022302462516e-16\\
43	1.11022302462516e-16\\
44	1.11022302462516e-16\\
45	1.11022302462516e-16\\
46	1.11022302462516e-16\\
47	1.11022302462516e-16\\
48	1.11022302462516e-16\\
49	1.11022302462516e-16\\
50	1.11022302462516e-16\\
};
\end{axis}

\end{tikzpicture}%
	\caption{The Euclidean norm of each sensor's estimation error, where the lines with `x' are those of misbehaving sensors.}
	\label{fig:err}
\end{figure}

The performance of Algorithm~\ref{alg:resilient} is demonstrated in Fig.~\ref{fig:err}, which shows the Euclidean norm of each sensor's estimation error. From the figure, we can see that, despite the presence of misbehaving nodes, the normal sensors can consistently estimate the true parameter, as expected from Theorem~\ref{thm:converge}.

\section{Conclusion}\label{sec:conclude}
This paper has considered the problem of distributed estimation in an adversarial environment, where some of the sensors might be Byzantine. To mitigate their misbehaviors, a resilient estimation algorithm has been proposed, which, with low computation complexity, guarantees that the normal sensors exponentially estimate the true parameter when certain requirements on the regressor vector and the network topology are met. As a future work, we plan to extend the results in this paper to more challenging scenarios where the parameter to be estimated might be changing dynamically and the sensor measurements might be subject to stochastic noises.

\bibliographystyle{IEEEtran}
\bibliography{reference}

\begin{thebibliography}{10}
\providecommand{\url}[1]{#1}
\csname url@samestyle\endcsname
\providecommand{\newblock}{\relax}
\providecommand{\bibinfo}[2]{#2}
\providecommand{\BIBentrySTDinterwordspacing}{\spaceskip=0pt\relax}
\providecommand{\BIBentryALTinterwordstretchfactor}{4}
\providecommand{\BIBentryALTinterwordspacing}{\spaceskip=\fontdimen2\font plus
\BIBentryALTinterwordstretchfactor\fontdimen3\font minus
  \fontdimen4\font\relax}
\providecommand{\BIBforeignlanguage}[2]{{%
\expandafter\ifx\csname l@#1\endcsname\relax
\typeout{** WARNING: IEEEtran.bst: No hyphenation pattern has been}%
\typeout{** loaded for the language `#1'. Using the pattern for}%
\typeout{** the default language instead.}%
\else
\language=\csname l@#1\endcsname
\fi
#2}}
\providecommand{\BIBdecl}{\relax}
\BIBdecl

\bibitem{goodwin2014adaptive}
G.~C. Goodwin and K.~S. Sin, \emph{Adaptive {F}iltering {P}rediction and
  {C}ontrol}.\hskip 1em plus 0.5em minus 0.4em\relax Courier, 2014.

\bibitem{xie2020convergence}
S.~Xie, Y.~Zhang, and L.~Guo, ``Convergence of a distributed least squares,''
  \emph{IEEE Transactions on Automatic Control}, vol.~66, no.~10, pp.
  4952--4959, 2020.

\bibitem{schizas2009distributed}
I.~D. Schizas, G.~Mateos, and G.~B. Giannakis, ``Distributed {LMS} for
  consensus-based in-network adaptive processing,'' \emph{IEEE Transactions on
  Signal Processing}, vol.~57, no.~6, pp. 2365--2382, 2009.

\bibitem{chen2013distributed}
W.~Chen, C.~Wen, S.~Hua, and C.~Sun, ``Distributed cooperative adaptive
  identification and control for a group of continuous-time systems with a
  cooperative {PE} condition via consensus,'' \emph{IEEE Transactions on
  Automatic Control}, vol.~59, no.~1, pp. 91--106, 2013.

\bibitem{yan2022distributed}
J.~Yan, X.~Yang, Y.~Mo, and K.~You, ``A distributed implementation of
  steady-state kalman filter,'' \emph{IEEE Transactions on Automatic Control},
  2022, DOI: 10.1109/TAC.2022.3175925.

\bibitem{xie2018analysis}
S.~Xie and L.~Guo, ``Analysis of distributed adaptive filters based on
  diffusion strategies over sensor networks,'' \emph{IEEE Transactions on
  Automatic Control}, vol.~63, no.~11, pp. 3643--3658, 2018.

\bibitem{matveev2021diffusion}
A.~S. Matveev, M.~Almodarresi, R.~Ortega, A.~Pyrkin, and S.~Xie,
  ``Diffusion-based distributed parameter estimation through directed graphs
  with switching topology: Application of dynamic regressor extension and
  mixing,'' \emph{IEEE Transactions on Automatic Control}, vol.~67, no.~8, pp.
  4256--4263, 2022.

\bibitem{chen2019resilient}
Y.~Chen, S.~Kar, and J.~M. Moura, ``Resilient distributed parameter estimation
  with heterogeneous data,'' \emph{IEEE Transactions on Signal Processing},
  vol.~67, no.~19, pp. 4918--4933, 2019.

\bibitem{mitra2019byzantine}
A.~Mitra and S.~Sundaram, ``{B}yzantine-resilient distributed observers for
  {LTI} systems,'' \emph{Automatica}, vol. 108, p. 108487, 2019.

\bibitem{dolev1986reaching}
D.~Dolev, N.~A. Lynch, S.~S. Pinter, E.~W. Stark, and W.~E. Weihl, ``Reaching
  approximate agreement in the presence of faults,'' \emph{Journal of the ACM
  (JACM)}, vol.~33, no.~3, pp. 499--516, 1986.

\bibitem{leblanc2013resilient}
H.~J. LeBlanc, H.~Zhang, X.~Koutsoukos, and S.~Sundaram, ``Resilient asymptotic
  consensus in robust networks,'' \emph{IEEE Journal on Selected Areas in
  Communications}, vol.~31, no.~4, pp. 766--781, 2013.

\bibitem{dibaji2017resilient}
S.~M. Dibaji, H.~Ishii, and R.~Tempo, ``Resilient randomized quantized
  consensus,'' \emph{IEEE Transactions on Automatic Control}, vol.~63, no.~8,
  pp. 2508--2522, 2017.

\bibitem{ishii2022overview}
H.~Ishii, Y.~Wang, and S.~Feng, ``An overview on multi-agent consensus under
  adversarial attacks,'' \emph{Annual Reviews in Control}, 2022.

\bibitem{leblanc2014resilient}
H.~J. LeBlanc and F.~Hassan, ``Resilient distributed parameter estimation in
  heterogeneous time-varying networks,'' in \emph{Proceedings of the 3rd
  International Conference on High Confidence Networked Systems}, 2014, pp.
  19--28.

\bibitem{meng2021distributed}
M.~Meng, X.~Li, and G.~Xiao, ``Distributed estimation under sensor attacks:
  Linear and nonlinear measurement models,'' \emph{IEEE Transactions on Signal
  and Information Processing over Networks}, vol.~7, pp. 156--165, 2021.

\bibitem{chen2018resilient}
Y.~Chen, S.~Kar, and J.~M. Moura, ``Resilient distributed estimation through
  adversary detection,'' \emph{IEEE Transactions on Signal Processing},
  vol.~66, no.~9, pp. 2455--2469, 2018.

\bibitem{aranovskiy2017performance}
S.~Aranovskiy, A.~Bobtsov, R.~Ortega, and A.~Pyrkin, ``Performance enhancement
  of parameter estimators via dynamic regressor extension and mixing,''
  \emph{IEEE Transactions on Automatic Control}, vol.~62, no.~7, pp.
  3546--3550, 2017.

\bibitem{mendes2013multidimensional}
H.~Mendes and M.~Herlihy, ``Multidimensional approximate agreement in
  {B}yzantine asynchronous systems,'' in \emph{Proceedings of the 45th Annual
  ACM Symposium on Theory of Computing}, 2013, pp. 391--400.

\bibitem{wang2018resilient}
X.~Wang, S.~Mou, and S.~Sundaram, ``A resilient convex combination for
  consensus-based distributed algorithms,'' \emph{arXiv preprint
  arXiv:1806.10271}, 2018.

\bibitem{yan2020resilient}
J.~Yan, X.~Li, Y.~Mo, and C.~Wen, ``Resilient multi-dimensional consensus in
  adversarial environment,'' \emph{Automatica}, vol. 145, p. 110530, 2022.

\bibitem{yi2022conditions}
B.~Yi and R.~Ortega, ``Conditions for convergence of dynamic regressor
  extension and mixing parameter estimators using lti filters,'' \emph{IEEE
  Transactions on Automatic Control}, 2022, DOI: 10.1109/TAC.2022.3149964.

\bibitem{vaidya2012iterative}
N.~H. Vaidya, L.~Tseng, and G.~Liang, ``Iterative approximate {B}yzantine
  consensus in arbitrary directed graphs,'' in \emph{Proceedings of the ACM
  Symposium on Principles of Distributed Computing}, 2012, pp. 365--374.

\bibitem{usevitch2019resilient}
J.~Usevitch and D.~Panagou, ``Resilient leader-follower consensus to arbitrary
  reference values in time-varying graphs,'' \emph{IEEE Transactions on
  Automatic Control}, vol.~65, no.~4, pp. 1755--1762, 2019.

\bibitem{yan2021resilient}
J.~Yan, C.~Deng, and C.~Wen, ``Resilient output regulation in heterogeneous
  networked systems under {B}yzantine agents,'' \emph{Automatica}, vol. 133, p.
  109872, 2021.

\end{thebibliography}
\end{document}